\documentclass{emulateapj}
\usepackage{amsmath}
\usepackage{amssymb}
\usepackage{amsthm}
\usepackage{array}
\usepackage{float}
\usepackage{graphics}
\usepackage{subfigure}

\newcommand{\pd}{\partial}  
\newcounter{ichi}
\newcounter{ni}
\newcounter{san}
\newcounter{yon}
\slugcomment{}

\shorttitle{}
\shortauthors{Murase et al.}

\begin{document}

\title{Gamma-Ray and Hard X-Ray Emission from Pulsar-Aided Supernovae\\
as a Probe of Particle Acceleration in Embryonic Pulsar Wind Nebulae}
\author{Kohta Murase\altaffilmark{1,2}, Kazumi Kashiyama\altaffilmark{2,3}, Kenta Kiuchi\altaffilmark{4}, and Imre Bartos\altaffilmark{5}}
\altaffiltext{1}{Hubble Fellow --- Institute for Advanced Study, Princeton, New Jersey 08540, USA}
\altaffiltext{2}{Center for Particle and Gravitational Astrophysics; Department of Physics; Department of Astronomy \& Astrophysics, Pennsylvania State University, University Park, Pennsylvania 16802, USA}
\altaffiltext{3}{Einstein Fellow --- Department of Astronomy and Theoretical Astrophysics Center; University of California, Berkeley, CA 94720, USA}
\altaffiltext{4}{Yukawa Institute for Theoretical Physics, Kyoto University, Kyoto, Kyoto 606-8502, Japan}
\altaffiltext{5}{Department of Physics; Columbia Astrophysics Laboratory, Columbia University, New York, New York 10027, USA}


\begin{abstract}
It has been suggested that some classes of luminous supernovae (SNe) and gamma-ray bursts (GRBs) are driven by newborn magnetars.  Fast-rotating proto-neutron stars have also been of interest as potential sources of gravitational waves (GWs).  
We show that for a range of rotation periods and magnetic fields, hard X rays and GeV gamma rays provide us with a promising probe of pulsar-aided SNe.  It is observationally known that young pulsar wind nebulae (PWNe) in the Milky Way are very efficient lepton accelerators.  We argue that, if embryonic PWNe satisfy similar conditions at early stages of SNe (in $\sim1-10$ months after the explosion), external inverse-Compton emission via upscatterings of SN photons is naturally expected in the GeV range as well as broadband synchrotron emission.  To fully take into account the Klein-Nishina effect and two-photon annihilation process that are important at early times, we perform detailed calculations including electromagnetic cascades.  
Our results suggest that hard X-ray telescopes such as {\it NuSTAR} can observe such early PWN emission by followup observations in months-to-years. GeV gamma rays may also be detected by \textit{Fermi} for nearby SNe, which serve as counterparts of these GW sources.  
Detecting the signals will give us an interesting probe of particle acceleration at early times of PWNe, as well as clues to driving mechanisms of luminous SNe and GRBs.  Since the Bethe-Heitler cross section is lower than the Thomson cross section, gamma rays would allow us to study subphotospheric dissipation.  We encourage searches for high-energy emission from nearby SNe, especially Type Ibc SNe including super-luminous objects. 
\end{abstract}


\section{Introduction}
Neutron stars (NSs) are left as remnants of core-collapse supernovae (SNe).  A proto-NS cools via radiation of MeV neutrinos in the Kelvin-Helmholtz timescale of $\sim10-100$~s, and the neutrino-driven wind forms a hot bubble in the SN cavity.  For fast-rotating NSs, their rotation energy is non-negligible and can be larger than the typical SN explosion energy $\sim{10}^{51}$~erg~\citep[e.g.,][]{og71}.  If NSs are magnetized and can supply enough plasma, it is extracted as the Poynting flux and the spin-down luminosity may contribute to observed SN emission.  Interestingly, recent surveys for optical and infrared transients have revealed a diversity of SN phenomena~\citep[e.g.,][]{arc+10,smi+11,gal12}, and some super-luminous SNe (especially hydrogen-poor SNe) and hypernovae can be explained as pulsar-powered SNe~\citep[e.g.,][]{whe+00,tho+04,kb10,woo10,met+11,nic+13,ins+13,mcc+14}. 
Even long gamma-ray bursts (GRBs) may be driven by fast-rotating magnetars with magnetic fields of $\sim{10}^{14}-{10}^{15}$~G if a proto-NS wind forms relativistic jets~\citep[e.g.,][]{uso92,by98,dl98,zm01,tho+04,mur+11grb,met+11}, and fast rotation with spin periods of $P_{i}\lesssim3$~ms has been expected in the dynamo scenario~\citep{dt92}. 

On the other hand, regardless of above theoretical expectations, pulsar wind nebulae (PWNe) with ages of $\gtrsim100$~yr have been observationally established as efficient particle accelerators~\citep{gs06}.  Measurements of the pulsar rotation period $P$ and its derivative $\dot{P}$ infers the spin-down luminosity.  In comparison with multi-wavelength observations, a significant fraction of the spin-down power is dissipated and used for acceleration of relativistic electrons and positrons~\citep{rg74,kc84fit,jag+96,aa96,tt10}.  Young PWNe such as the Crab nebula suggest that accelerated leptons typically have TeV energies and radiate synchrotron and inverse-Compton (IC) photons at broad wavelengths.  However, particle acceleration mechanisms before or around the termination shock are not well-known.  In the classical model, the magnetic energy needs to be converted into the kinetic energy~\citep{rg74,kc84fit}, but details of magnetic dissipation and related processes such as the kink instability are still under debate~\citep[e.g.,][]{beg98,lk01,aro12,por+14,cer+14}.   

In this work, we consider implications of such PWNe for early SN emission, focusing on hard X rays and gamma-ray emission.  If embryonic PWNe are efficient lepton accelerators as seen in Galactic PWNe, broadband non-thermal emission should naturally be expected as well~\citep[e.g.,][]{vp78,gs87,vp05,mp13,kot+13}.  In particular, \cite{kot+13} recently argued that TeV gamma rays provide promising signals that are detectable by gamma-ray telescopes.  Future gamma-ray telescopes such as the Cherenkov Telescope Array (CTA)~\citep{cta+11} will also be useful.  Another interesting signature is predicted at soft X rays~\citep{met+14}, and breakout gamma-ray emission is also expected a bit earlier than breakout optical emission.  At early stages, the photon density in a wind bubble are large enough for the two-photon annihilation process and subsequent electromagnetic cascades should occur.  The surrounding ejecta density is also quite large, where X rays and gamma rays are significantly attenuated.  Focusing on hard X rays and gamma rays, we here provide detailed studies of high-energy photon spectra of pulsar-aided SNe~\footnote{Here we consider both cases where spin-down power is dominant (i.e., pulsar-powered SNe) and sub-dominant.}, and show that their non-thermal signatures are useful to understand how efficient lepton acceleration begins in PWNe.  Driving mechanisms of super-luminous SNe, hypernovae and GRBs are unknown.  Thus, successful detections of such high-energy signals, which support the existence of fast-rotating NSs embedded in stellar material, will help us reveal links between these energetic SNe and ordinary SNe.  In addition, proto-NSs have also been considered as promising sources of gravitational waves (GWs)~\citep[see reviews, e.g.,][and references therein]{kok08,bar+13}.  In particular, if a millisecond pulsar is significantly deformed, a fraction of the rotational energy can be emitted as GWs~\citep[e.g.,][]{og69,cut02,ste+05,dal+09}, which is detectable by second-generation ground-based GW interferometers such as Advanced LIGO~\citep[][]{aligo}, Advanced Virgo~\citep[][]{avirgo}, and KAGRA~\citep[][]{kagra}.  

In Section~2, we describe the basic picture and method of calculation.  We also provide analytic spectra, taking into account the Klein-Nishina (KN) effect that is relevant in our problem.  We show our numerical results in Section~3.  In Section~4, we additionally discuss related issues and then summarize our results.

Throughout this work, we use the notation $Q={10}^xQ_x$ in CGS unit unless noted otherwise.

\section{Basic Setup}
\subsection{Dynamics}
A massive star with $\gtrsim8 M_{\odot}$ has been believed to cause a SN explosion, leaving a proto-NS or black hole.  The NS is initially hot, and cools down in the Kevin-Helmholtz time scale of $\sim10-100$~s.  Initially, mass losses are mainly caused by a thermal neutrino-driven wind, and a hot wind-driven bubble forms in the SN cavity.  When the NS is rotating and magnetized, its early pulsar wind is expected to become Poynting-dominated and relativistic~\citep[e.g.,][]{tho+04}.  Then, as in Galactic PWNe, one may expect that almost all the spin-down power is converted into radiation.  The rotation energy budget is 
\begin{equation}\label{eq:rot}
{\mathcal E}_{{\rm rot},i} =  \frac{I (2\pi/P_i)^2}{2} \simeq 2.8 \times{10}^{51}~{\rm erg}~P_{i,-2.5}^{-2},
\end{equation}
where $I\approx0.35 M_{\rm ns} R_{\rm ns}^2 \simeq1.4\times10^{45}~\rm g~cm^2$ is the momentum of inertia~\citep{lp01}, where $M_{\rm ns} = 1.4~M_\odot$ and $R_{\rm ns} = 12~{\rm km}$ are used.  
The initial mass-loss rate is governed by the neutrino-driven wind, and then the wind is carried by electrons and/or positrons especially after the proto-NS is transparent to neutrinos.  In the late phase, the spin down of the NS is approximated by
\begin{equation}\label{eq:L}
-\frac{d {\mathcal E}_{\rm rot}}{dt} = L_{\rm em} + L_{\rm gw},
\end{equation}
where the electromagnetic spin-down luminosity is
\begin{eqnarray}\label{eq:L_m_1}
L_{\rm em}&\approx&\frac{\mu^2 (2\pi/P)^4}{c^3} (1+ C\sin^2 \chi_\mu).
\end{eqnarray}
Here $\mu\equiv0.5B_{\rm dip}R_{\rm ns}^3$ is the magnetic moment, $C\sim1$ is a pre-factor suggested from magnetohydrodynamics simulations~\citep{gru05,spi06,tch+13}, and $\chi_\mu$ is the angle between the magnetic and rotation axes.  Rotating proto-NSs can be unstable to non-axisymmetric deformations, potentially causing strong GW emission via rotation instabilities including dynamical or secular one, and/or via magnetic distortion~\citep[see][and references therein]{kok08,cm09,bar+13}.  For instance, in the quadrupole approximation, the GW luminosity can roughly be estimated to be
\begin{equation}\label{eq:L_gw_1}
L_{\rm gw}\sim\frac{32}{5}\frac{G(\epsilon_G I)^2 (2 \pi/P')^6}{c^5}, 
\end{equation}
$\epsilon_G$ is the ellipticity and $P'$ is the pattern period of the elliptical figure. 
In particular, strong toroidal magnetic fields may make the NS prolate, and the configuration can increase the angle between deformation and rotation axes until they are orthogonal, where the GW emission can be described by the quadrupole emission of a rotating, non-axisymmetric body deformed by internal magnetic fields~\citep[e.g.,][]{cut02,ste+05,dal+09}.  

In most cases in which we are interested, $L_{\rm em}$ is dominant, and $L_{\rm em}$ is estimated to be~\footnote{Different expressions of $L_{\rm em}$ lead to different numerical values.  When we adopt the conventional magnetic dipole formula, we have $L_{\rm em}=\frac{4}{9}\frac{\mu^2(2\pi/P)^4}{c^3}$~\citep{og69}.  In this case, at $t\gg t_{\rm em}$, we have $L_{\rm em}\simeq6.1\times{10}^{47}~{\rm erg}~{\rm s}^{-1}~B_{\rm dip,15}^{-2}I_{45}^2R_{\rm ns,6}^{-6}t_{4}^{-2}$ or $4.0\times{10}^{43}~{\rm erg}~{\rm s}^{-1}~B_{\rm dip,14}^{-2}t_{7}^{-2}$ for $M_{\rm ns} = 1.4~M_\odot$ and $R_{\rm ns} = 12~{\rm km}$~\citep{mur+09}.  Note that magnetic dissipation in the current sheet may reduce $L_{\rm em}$.} 
\begin{eqnarray}\label{eq:L_m_2}
L_{\rm em}&=&L_{{\rm em},i}{\left(1+\frac{t}{t_{\rm em}}\right)}^{-2}\nonumber\\
&\simeq&\left\{ \begin{array}{ll}
8.6\times{10}^{45}~{\rm erg}~{\rm s}^{-1}~P_{i,-2.5}^{-4}B_{\rm dip,14}^2
& \mbox{($t\leq t_{\rm em}$)}\\
8.9\times{10}^{42}~{\rm erg}~{\rm s}^{-1}~B_{\rm dip,14}^{-2}t_{7}^{-2}
& \mbox{($t>t_{\rm em}$)}
\end{array} \right. 
\end{eqnarray}
Here the characteristic spin-down time is given by
\begin{equation}
t_{\rm em}=\frac{P_i^2Ic^3}{4\pi^2B_{\rm dip,14}^2R_{\rm ns}^6}\simeq3.2\times{10}^5~{\rm s}~B_{\rm dip,14}^{-2}P_{i,-2.5}^{2}.
\end{equation}

The fate of the early PWN depends on various parameters such as the spin-down power and baryon loading.  If the spin-down power is high enough, some two-dimensional simulations suggest that the equatorial wind can be redirected by the anisotropic pressure, and hoop stresses lead to bipolar outflows~\footnote{In this case, the (collimated) wind radius is $R_w\approx ct$.} that could explain GRBs~\citep{buc+07,kb07,buc+08}.  If not, we expect a quasi-spherical expanding flow embedded in the expanding stellar material (see Figure~1).  Assuming a SN explosion with ${\mathcal E}_{\rm sn}\sim{10}^{51}$~erg, the SN ejecta expands with its velocity $V_{\rm ej}$ and radius $R_{\rm ej}$.  The early PWN radius $R_{w}$ also increases non-relativistically, which is given by~\citep[e.g.,][]{met+14}
\begin{equation}\label{eq:nb}
\frac{d R_{w}}{dt}=\sqrt{\frac{7}{6(3-\delta)}\frac{{\mathcal E}_{\rm rot}}{M_{\rm ej}}{\left(\frac{R_w}{R_{\rm ej}}\right)}^{3-\delta}}+\frac{R_{w}}{t},
\end{equation}
for $R_{w}<R_{\rm ej}$, otherwise $R_{w}\approx R_{\rm ej}$ is used.  Note that we have used the ejecta density
\begin{equation}
\rho_{\rm ej}=\frac{(3-\delta)M_{\rm ej}}{4\pi R_{\rm ej}^3}{\left(\frac{R}{R_{\rm ej}}\right)}^{-\delta},
\end{equation}
where $\delta\sim0-1$ is a typical value used in the literature~\citep{kb10,met+14}.  The mixture of material allows us to approximate the inner density profile to be reasonably smooth and flat~\citep{che77,cf92}.  For demonstration, we adopt $\delta=1$ throughout this work~\citep{kb10,met+14}, and that the radiation pressure is given by ${\mathcal E}_{\rm rot}/(3{\mathcal V}_{\rm nb})\approx(6/7)\rho_{\rm ej}V_{\rm nb}^2$.  Here ${\mathcal V}_{\rm nb}$ is the PWN volume and $V_{\rm nb}$ is the PWN expansion velocity that can be different from $V_{\rm ej}$.  In general, $R_w$ is smaller than $R_{\rm ej}$, and both of $R_{\rm ej}$ and $R_w$ are numerically determined in this work.  Roughly speaking, $R_{w}\approx R_{\rm ej}$ becomes a good approximation for small values of $P$ such that ${\mathcal E}_{{\rm rot},i}\gtrsim{\mathcal E}_{\rm sn}$ (implying $P_{i}\lesssim5~{\rm ms}~{\mathcal E}_{\rm sn,51}^{-1/2}$).  The ejecta velocity $V_{\rm ej}$ and radius $R_{\rm ej}$ can be determined by
\begin{equation}
V_{\rm ej}=\sqrt{\frac{2[\int dt \,\, ({\mathcal E}_{\rm int}/t_{\rm dyn})+{\mathcal E}_{\rm sn}]}{M_{\rm ej}}}
\end{equation}
\begin{equation}\label{eq:ej}
\frac{d R_{\rm ej}}{dt}=V_{\rm ej}. 
\end{equation}
The internal energy trapped in the SN ejecta, ${\mathcal E}_{\rm int}$, is given by
\begin{equation}\label{eq:int}
\frac{d {\mathcal E}_{\rm int}}{dt} = L_{\rm em}- \frac{{\mathcal E}_{\rm int}}{t_{\rm dyn}} - \frac{{\mathcal E}_{\rm int}}{t_{\rm esc}^{\rm ej}},
\end{equation}
where $t_{\rm dyn}\approx R_{\rm ej}/V_{\rm ej}$ is the dynamical time.  
Since X-ray and gamma-ray emission is expected in month-to-year time scales, we only consider energy injection due to $L_{\rm em}$.  In the early phase, as in normal SNe, heating by shocks and unstable isotopes such as ${}^{56}$Ni can be relevant.  In the later phase, one may assume that late interactions with circumstellar material are negligible, and injections via the $\beta$ decay of ${}^{56}$Ni are irrelevant after their lifetime $t_{{}^{56}{\rm Ni}}=6.075~{\rm d}\simeq5.2\times10^5~{\rm s}$.
Visible photons leave the SN ejecta in the escape time
\begin{equation}
t_{\rm esc}^{\rm ej}\approx\frac{(1+\tau_T^{\rm ej})R_{\rm ej}}{c},   
\end{equation}
where the Thomson optical depth in the ejecta is given by $\tau_T^{\rm ej}\approx K_{T}\rho_{\rm ej}(R_{\rm ej})R_{\rm ej}$, which is estimated to be
\begin{eqnarray}\label{eq:Thomson}
\tau_T^{\rm ej}\approx\frac{(3-\delta)K_{T}M_{\rm ej}}{4\pi R_{\rm ej}^2}&\simeq&13~(2/\mu_e)(M_{\rm ej}/5~M_\odot)\nonumber\\
&\times&{(V_{\rm ej}/5000~{\rm km}~{\rm s}^{-1})}^{-2}t_7^{-2},
\end{eqnarray}
where $K_T=\mu_e^{-1}\sigma_T/m_u$, $\mu_e$ is the mean molecular weight per electron, and $m_u$ is the atomic mass unit.  See also Equation~(45) below.  The two of the key parameters, $E_{\rm sn}$ and $M_{\rm ej}$, can be estimated from the SN peak emission and determination of the ejecta velocity $V_{\rm ej}$ via detailed spectroscopy.  Note that the bound-free or bound-bound cross section is much higher at $\lesssim10$~keV energies, and thermal photons are still generated at later times.    

Non-thermal photons generated in the PWN are significantly thermalized in the SN ejecta.  Since we are interested in the IC emission, we need to estimate a thermal component, which serves as a seed photon field.  Ideally, self-consistent calculations including the detailed radiative transfer are needed.  But, for the present purpose, the following approximate approach is sufficient.  The internal energy is divided into the thermal energy ${\mathcal E}_{\rm th}$ and non-thermal energy ${\mathcal E}_{\rm nonth}$.  Following \cite{kas+14}, the thermal energy is calculated by
\begin{equation}\label{eq:th}
\frac{d {\mathcal E}_{\rm th}}{dt} =  \int dE_\gamma \,\, \frac{(1-{\mathcal A}_{E_\gamma}) E_\gamma {\mathcal N}_{E_\gamma}}{t_{\rm esc}^{\rm ej}} - \frac{{\mathcal E}_{\rm th}}{t_{\rm dyn}} - \frac{{\mathcal E}_{\rm th}}{t_{\rm esc}^{\rm ej}},
\end{equation}
where ${\mathcal N}_{E_\gamma}$ is the differential photon number (per energy) and ${\mathcal A}_{E_\gamma}$ is the energy-dependent albedo factor, i.e., the fraction of photons escaping without thermalization.  In this work, for simplicity, we use ${\mathcal A}_{E_\gamma}=0.5$ for photon energies below the cutoff due to Compton down-scattering in the SN ejecta, otherwise we set ${\mathcal A}_{E_\gamma}=0$.  Because of the photoelectric absorption (see Section~2.4), soft X rays and UV photons may not escape until very late times, so our choice is reasonable.  Lower values simply imply that more energy is thermalized, leading to brighter SN emission.  Also, in the pulsar-aided SN model, ${\mathcal A}_{E_\gamma}$ can be phenomenologically adjusted to explain observed SN emission~\citep{kas+14}.  
Then, the SN temperature is approximated to be $T_{\rm sn}={[{\mathcal E}_{\rm th}/(a{\mathcal V}_{\rm ej})]}^{1/4}$, which also gives the photon density of target photons.  Here $a$ is the Stefan-Boltzmann constant and ${\mathcal V}_{\rm ej}$ is the volume of the SN ejecta.

Note that interactions in the PWN are negligible in our setup.  Assuming the pair multiplicity $\mu_\pm$ is larger than $m_p/m_e$, the Thomson optical depth in the PWN ($\tau_T^{\rm nb}\approx\sigma_Tn_{\rm nb}R_w$) is estimated to be~\footnote{Note that the Thomson optical depth in the relativistic wind is smaller by a factor of $\Gamma_w^2$, so thermalization there is relevant only in the earliest phase.} 
\begin{eqnarray}
\tau_T^{\rm nb}\approx\frac{\sigma_T\dot{M}_{\pm}}{4\pi R_{w}V_{\rm nb}m_e}
&\simeq&2.0\times{10}^{-6}~P_{-2.5}^{-2}B_{\rm dip,14}\nonumber\\
&\times&{\left(\frac{V_{\rm nb}}{5000~{\rm km}~{\rm s}^{-1}}\right)}^{-2}t_7^{-1}\mu_{\pm,6},
\end{eqnarray} 
where the nebula density $n_{\rm nb}\approx\dot{M}_{\pm}/(4\pi R_w^2V_{\rm nb}m_e)$ is used and $\dot{M}_{\pm}$ is the mass-loss rate according to the Goldreich-Julian density multiplied by $\mu_\pm$~\citep{gj69}, and $P$ is a function of $t$.

\begin{figure}
\includegraphics[width=\linewidth]{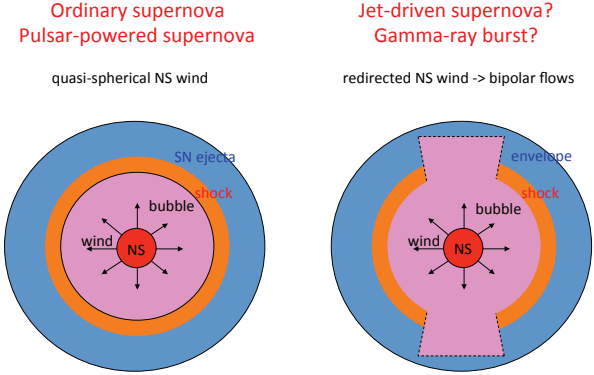}
\caption{The schematic picture of pulsar-aided SNe.  We consider the left case, where a pulsar wind is quasi-spherical and the wind bubble is embedded in the SN ejecta.  
}
\end{figure}

\subsection{Leptonic Emission from Embryonic PWNe}
It has been known that Galactic PWNe are efficient accelerators of electrons and positrons.  The Crab pulsar is one of the most well-known high-energy gamma-ray sources, and $P_i\approx19$~ms and $B_{\rm dip}\approx5\times{10}^{12}$~G are indicated~\citep[e.g.,][]{fk06}.  We assume that lepton acceleration can occur even in the very early stage of PWNe similarly to that in Galactic PWNe, and that some extragalactic SNe leave fast-rotating ($P\lesssim10$~ms) and strongly magnetized ($B_{\rm dip}\gtrsim{10}^{13}$~G) NSs.  For relativistic leptons, we consider a broken power-law injection spectrum, 
\begin{equation}\label{eq:inj}
\frac{d \dot{\mathcal N}_e^{\rm inj}}{d\gamma_e}
\propto
\left\{ \begin{array}{ll} 
\gamma_e^{-q_1}
& \mbox{($\gamma_m\leq\gamma_e\leq\gamma_b$)}\\
\gamma_e^{-q_2}
& \mbox{($\gamma_b<\gamma_e\leq\gamma_M$)}
\end{array} \right.
\end{equation} 
where $q_1(<2)$ and $q_2(>2)$ are low- and high-energy injection spectral indices, and $\gamma_b$ is the break Lorentz factor.  Fitting results on some Galactic PWNe suggest $q_1\sim1-1.5$, $q_2\sim2.5-3$, $\gamma_b={10}^{4.5}-{10}^6$, and the significant energy fraction $\epsilon_e\sim1$ of $L_{\rm em}$ is carried by leptons~\citep{tt10,tt13}.  The maximum Lorentz factor of accelerated electrons and positrons, $\gamma_M$, is given by the balance between the acceleration time and cooling time (see below).  The minimum injection Lorentz factor is assumed to be $\gamma_m=100$ but our results are insensitive to it as long as its value is small enough.  In this work, we expect that the wind is largely dominated by pairs so ion acceleration is negligible, where $\mu_\pm$ and $\gamma_b$ can be related~\footnote{The pair multiplicity at $t\gg t_{\rm em}$ is not far from the values obtained for Galactic PWNe.  If $\gamma_b>\gamma_M$, acceleration of pairs is limited by strong radiative cooling.} as $\mu_\pm\sim{10}^9~\epsilon_e\gamma_{b,5}^{-1}{(\gamma_b/\gamma_m)}^{q_1-1}[(2-q_1)(q_2-2)/(q_1-1)/(q_2-q_1)]B_{\rm dip,14}P_{-2.5}^{-2}$ from number and energy conservation.  Details would depend on physics of potential drops and dissipation in the current sheet, where pair production with external photons provided by SN emission plays a role.  Note that, for the bulk Lorentz factor $\Gamma_w$ and the magnetization parameter $\sigma$, one also has $\mu_{\pm}\simeq1.2\times{10}^{14}{[\Gamma_w(1+\sigma)]}^{-1}L_{\rm em,46}^{1/2}$.

Leptons rapidly cool via synchrotron and IC emission mechanisms.  In this work, the magnetic field energy density in the early PWN is given by
\begin{equation}
U_B=\epsilon_B \frac{3\int dt\,\,L_{\rm em}}{4\pi R_{w}^3},
\end{equation}
where $\epsilon_B={10}^{-3}-{10}^{-2}$ are indicated~\citep[e.g.,][]{kc84fit,jag+96,aa96,tt10}.  The magnetic field is estimated to be
\begin{eqnarray}
B&\simeq&36~{\rm G}~P_{i,-2.5}^{-1}\epsilon_{B,-2}^{1/2}{\left(\frac{V_{\rm ej}}{5000~{\rm km}~{\rm s}^{-1}}\right)}^{-3/2}\nonumber\\
&\times&t_7^{-3/2}{\left[1-{\left(1+t/t_{\rm em}\right)}^{-1}\right]}^{1/2},
\end{eqnarray}
where $R_w=R_{\rm ej}=V_{\rm ej}t$ is used for analytical estimates~\footnote{In our numerical calculations, $R_w$, $R_{\rm ej}$ and $V_{\rm ej}$ are obtained by solving differential equations.}.  
In Figure~2, we plot synchrotron and IC cooling time scales, as well as the dynamical time $t_{\rm dyn}\approx R_{\rm ej}/V_{\rm ej}$.  One immediately sees the energy dependence of the synchrotron cooling time $t_{\rm syn}\approx 3m_ec/(4\sigma_TU_B\gamma_e)$, whereas the IC cooling time deviates from the expectation in the Thomson regime, $t_{\rm IC}\propto \gamma_e^{-1}$, due to the KN effect.

The radiative cooling time scale is given by $t_{\rm rad}^{-1}=t_{\rm syn}^{-1}+t_{\rm IC}^{-1}=t_{\rm syn}^{-1}(1+Y)$, where $Y=t_{\rm syn}/t_{\rm IC}$ is the total Compton $Y$ parameter.  Then, at $t\gg t_{\rm em}$, the cooling Lorentz factor of electrons is estimated to be
\begin{equation}
\gamma_{c} \simeq1.9\times{10}^{-2}P_{i,-2.5}^{2}\epsilon_{B,-2}^{-1}{\left(\frac{V_{\rm ej}}{5000~{\rm km}~{\rm s}^{-1}}\right)}^{3}t_7^2{(1+Y)}^{-1},
\end{equation}
where $t_{\rm rad}=t_{\rm dyn}$ is used.  One should keep in mind $\gamma_e$ cannot be less than unity physically.  If $\gamma_c<1$ in Equation~(19), it simply implies that relativistic electrons will become non-relativistic in $t_{\rm dyn}$ due to strong cooling.  
Note that, in the Thomson limit, the $Y$ parameter is roughly given by
\begin{equation}
Y\approx\frac{-1+\frac{L_{\rm sn}tV_{\rm ej}}{\epsilon_B {\mathcal E}_{\rm em}c}+\sqrt{{\left(1+\frac{L_{\rm sn}tV_{\rm ej}}{\epsilon_B {\mathcal E}_{\rm em}c}\right)}^2+\frac{4\epsilon_e L_{\rm em}tV_{\rm ej}}{\epsilon_B {\mathcal E}_{\rm em}c}}}{2} 
\end{equation}
The distribution of pairs is essentially in the fast cooling regime.  In the fast cooling case ($\gamma_{c} < \gamma_{m}$) with constant $Y$, the steady-state electron distribution is $d \mathcal{N}_e/d \gamma_e \propto \gamma_e^{-2}$ for $1\lesssim\gamma_e\leq\gamma_{m}$, $d \mathcal{N}_e/d \gamma_e \propto \gamma_e^{-q_1-1}$ for $\gamma_{m}\leq\gamma_e\leq\gamma_{b}$ and $d\mathcal{N}_e/d \gamma_e\propto\gamma_e^{-q_2-1}$ for $\gamma_{b}\leq\gamma_e\leq\gamma_M$.  
Using the particle acceleration time $t_{\rm acc}=\eta r_L/c=\eta \gamma_e m_ec^2/eBc$ (where $\eta\geq1$ is the pre-factor accounting for acceleration efficiency), the pair Lorentz factor is limited by
\begin{eqnarray}
\gamma_M\approx\sqrt{\frac{6\pi e\eta^{-1}}{\sigma_TB (1+Y_M)}}&\simeq&1.9\times{10}^{7}~P_{i,-2.5}^{1/2}\eta^{-1/2}\epsilon_{B,-2}^{-1/4}\nonumber\\
&\times&{\left(\frac{V_{\rm ej}}{5000~{\rm km}~{\rm s}^{-1}}\right)}^{3/4}\frac{t_7^{3/4}}{{(1+Y_M)}^{1/2}},\,\,\,\,\,\,\,\,\,\,
\end{eqnarray}
which is obtained from $t_{\rm acc}=t_{\rm rad}$, and $Y_M\equiv Y(\gamma_M)$ is used.  The gamma-ray energy should be lower than
\begin{eqnarray}\label{eq:ICmax}
E_\gamma^M\approx\gamma_M m_ec^2&\simeq&9.9~{\rm TeV}~P_{i,-2.5}^{1/2}\eta^{-1/2}\epsilon_{B,-2}^{-1/4}\nonumber\\
&\times&{\left(\frac{V_{\rm ej}}{5000~{\rm km}~{\rm s}^{-1}}\right)}^{3/4}\frac{t_7^{3/4}}{{(1+Y_M)}^{1/2}}, 
\end{eqnarray} 
implying that $\gtrsim10-100$~TeV gamma rays are not expected at early stages of the PWN. 

\begin{figure}
\includegraphics[width=\linewidth]{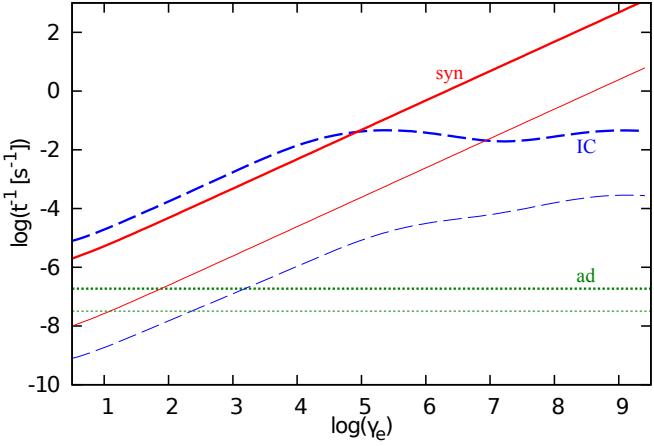}
\caption{Cooling timescales of electrons and positrons for ($P_{i}$, $B_{\rm dip}$, $M_{\rm ej}$)=(2~ms, ${10}^{14}$~G, 5~$M_{\odot}$) at $t={10}^{6.75}~{\rm s}~\simeq65$~d (thick curves) and $t={10}^{7.5}~{\rm s}~\simeq366$~d (thin curves).  One can see that relativistic pairs are in the fast cooling regime.   
}
\end{figure}

In the fast cooling case, the synchrotron photon spectrum is given by 
\begin{equation}
E_\gamma L_{E_\gamma}^{\rm syn}
\sim\frac{\epsilon_eL_{\rm em}}{2(1+Y){\mathcal R}_b}
\left\{ \begin{array}{ll} 
{(E_\gamma/E_{\rm syn}^b)}^{(2-q_1)/2}
& \mbox{($E_\gamma \leq E_{\rm syn}^b$)}\\
{(E_\gamma/E_{\rm syn}^b)}^{(2-q_2)/2}
& \mbox{($E_{\rm syn}^b \leq E_\gamma$)}
\end{array} \right.
\end{equation}
where ${\mathcal R}_b\sim{(2-q_1)}^{-1}+{(q_2-2)}^{-1}$.  The characteristic synchrotron energy is given by
\begin{eqnarray}
E_{\rm syn}^b\approx\frac{3}{2}\hbar\gamma_b^2\frac{eB}{m_ec}&\simeq&6.3~{\rm keV}~\gamma_{b,5}^2P_{i,-2.5}^{-1}\epsilon_{B,-2}^{1/2}\nonumber\\
&\times&{\left(\frac{V_{\rm ej}}{5000~{\rm km}~{\rm s}^{-1}}\right)}^{-3/2}t_7^{-3/2},
\end{eqnarray}
so the peak energy is expected in the X-ray range.  Note that the synchrotron maximum energy is
\begin{eqnarray}\label{eq:syncut}
E_{\rm syn}^M\approx\frac{3}{2}\hbar\gamma_M^2\frac{eB}{m_ec}&\approx&\hbar\frac{9\pi e^2}{\eta\sigma_Tm_e c(1+Y_M)}\nonumber\\
&\simeq&240~{\rm MeV}~\eta^{-1}{(1+Y_M)}^{-1},
\end{eqnarray}
which hardly depends on various parameters. 

The basic process for high-energy gamma-ray emission is the IC mechanism.  The expected IC luminosity~\footnote{If $\mathcal{Y}$ is introduced as the ratio of the IC energy flux to the seed photon energy flux, one may write $L_{\rm IC}\sim{\rm min} (\mathcal{Y}_{\rm syn} L_{\rm syn}, L_e)$.} is very roughly written as $L_{\rm IC}\sim Y{(1+Y)}^{-1}L_e$.  
First, let us assume that the seed photon spectrum has $E_\gamma L_{E_\gamma}\propto E_\gamma^{2-\beta}$ with $\beta\leq1+q_1/2$. Note that the synchrotron self-Compton (SSC) case corresponds to $\beta=1+q_1/2$ (in the fast cooling regime).  Then, the IC photon spectrum in the Thomson limit is expressed to be
\begin{equation}
E_\gamma L_{E_\gamma}^{\rm IC}
\propto
\left\{ \begin{array}{ll} 
E_\gamma^{(2-q_1)/2}
& \mbox{($E_\gamma\leq E_{\rm IC}^b$)}\\
E_\gamma^{(2-q_2)/2}
& \mbox{($E_{\rm IC}^b<E_\gamma$)}
\end{array} \right.
\end{equation}
One can obtain the above expression, noting that $L_{E_\gamma}^{\rm IC}\sim\int d \gamma_e (d \tau_{\rm IC}/d \gamma_e) L_{E}^{\rm seed} (\gamma_e,E)$, where $\gamma_e (d \tau_{\rm IC}/d \gamma_e)\propto \gamma_e^{-q_1}$ for $\gamma_e\leq\gamma_{b}$ and $\gamma_e (d \tau_{\rm IC}/d \gamma_e)\propto \gamma_e^{-q_2}$ for $\gamma_b<\gamma_{e}$, and $\tau_{\rm IC}$ is the IC optical depth.  
The similar spectrum is expected in the Thomson limit, when the seed photon spectrum is thermal.  
In the SSC case, the typical IC energy is
\begin{eqnarray}
E_{\rm SSC}^b\approx2\gamma_b^2E_{\rm syn}^b&\simeq&130~{\rm TeV}~\gamma_{b,5}^4P_{i,-2.5}^{-1}\epsilon_{B,-2}^{1/2}\nonumber \\
&\times&{\left(\frac{V_{\rm ej}}{5000~{\rm km}~{\rm s}^{-1}}\right)}^{-3/2}t_7^{-3/2},
\end{eqnarray}
but such high energies are difficult to achieve at early times due to Equation~(\ref{eq:ICmax}).  
In days-to-months time scales, SN emission can be prominent, where thermal photons are upscattered by relativistic pairs via the external IC (EIC) process.  The energy flux of seed photons has a peak at $E_{\rm sn}\approx3.92kT_{\rm sn}$, and the typical IC energy is
\begin{eqnarray}
E_{\rm EIC}^b\approx2\gamma_b^2E_{\rm sn}&\simeq&78~{\rm GeV}~\gamma_{b,5}^2(kT_{\rm sn}/1~{\rm eV}).
\end{eqnarray}
Note that Equation~(26) can be used for both SSC and EIC cases. 

However, the KN effect becomes very important at sufficiently high energies.  Let us introduce two characteristic energies~\citep{mur+11grb},
\begin{eqnarray}
E_{\rm KN}^{\rm typ}&\approx&m_e^2 c^4/(2E_{\rm typ}),\\
E_{\rm KN}^{b}&\approx&\gamma_{b}m_ec^2,
\end{eqnarray}
where $E_{\rm typ}$ is the typical energy of target photons.  In the SSC and EIC cases, we expect $E_{\rm typ}\approx E_{\rm syn}^b$ and $E_{\rm typ}\approx E_{\rm sn}$, respectively.  In the presence of the KN effect, IC spectra become complicated so we take a numerical approach (see Section~3).  However, it is useful to see analytical expressions.  First, we consider a seed photon spectrum of $E_\gamma L_{E_\gamma}\propto E_\gamma^{2-\beta}$.  Introducing $E_{\rm KN,1}$ as the first break energy due to the KN effect, for $E_{\rm KN,1}>E_{\rm IC}^b$, we have~\citep[e.g.,][]{mur+10,mur+11grb} 
\begin{equation}
E_\gamma L_{E_\gamma}^{\rm IC} 
\propto
\left\{ \begin{array}{ll} 
E_\gamma^{(2-q_1)/2}
& \mbox{($E_\gamma\leq E_{\rm IC}^b$)}\\
E_\gamma^{(2-q_2)/2}
& \mbox{($E_{\rm IC}^b<E_\gamma \leq E_{\rm KN,1}$)}\\
E_\gamma^{\beta-q_2}
& \mbox{($E_{\rm KN,1}\leq E_\gamma$)}
\end{array} \right.
\end{equation}
where the first KN break is given by
\begin{equation}
E_{\rm KN,1}=E_{\rm KN}^{\rm typ}\simeq33~{\rm GeV}~{(E_{\rm typ}/4~{\rm eV})}^{-1}.
\end{equation}
The IC emission at $E_\gamma>E_{\rm KN,1}$ is dominated by Thomson scattering between pairs with $\gamma_e\sim E_\gamma/(m_e c^2)$ and seed photons with $E\sim m_e^2 c^4/(2E_\gamma)$. 

If the first KN break appears below $E_{\rm IC}^b$, we obtain~\footnote{Equation~(23) in Murase et al. (2011b) corresponds to Equation~(\ref{eq:SSCtyp}), so $\beta_l=1+(p-1)/2$ is assumed.}  
\begin{equation}
E_\gamma L_{E_\gamma}^{\rm IC} 
\propto
\left\{ \begin{array}{ll} 
E_\gamma^{(2-q_1)/2}
& \mbox{($E_\gamma\leq E_{\rm KN,1}$)}\\
E_\gamma^{\beta-q_1}
& \mbox{($E_{\rm KN,1}<E_\gamma\leq E_{\rm KN,2}$)}\\
E_\gamma^{\beta-q_2}
& \mbox{($E_{\rm KN,2}\leq E_\gamma$)}
\end{array} \right.
\end{equation}
where $E_{\rm KN,1}=E_{\rm KN}^{\rm typ}$ and 
\begin{equation}
E_{\rm KN,2}=E_{\rm KN}^b\simeq51~{\rm GeV}~\gamma_{b,5}.
\end{equation}
is the second KN break.  Note that, if $\beta=1+q_1/2$ (as expected in the SSC case), we have 
\begin{equation}\label{eq:SSCtyp}
E_\gamma L_{E_\gamma}^{\rm IC} 
\propto
\left\{ \begin{array}{ll} 
E_\gamma^{(2-q_1)/2}
& \mbox{($E_\gamma\leq E_{\rm KN,1}$)}\\
E_\gamma^{\beta-q_2}
& \mbox{($E_{\rm KN,1}\leq E_\gamma$)}
\end{array} \right.
\end{equation}
where $E_{\rm KN,1}=E_{\rm KN}^{b}$.  This spectrum can be realized in SSC emission from the early PWN, but the break at $E_{\rm KN,1}$ is significantly smeared out because leptons upscattering photons with $E_{\rm typ}$ do not contribute above $E_{\rm KN}^{\rm typ}$.  
We do not consider cases where $\gamma_m$ and $\gamma_M$ enter expressions, since we assume that $\gamma_m$ is sufficiently small and $\gamma_M$ is sufficiently large. 

In our setup, EIC emission due to SN photons is often more important for gamma-ray detections.  When the seed photon spectrum is thermal, because the Rayleigh-Jeans spectrum is quite hard, the KN cross section becomes essential.  For $E_{\rm KN,1}>E_{\rm IC}^b$, we expect
\begin{equation}\label{eq:EICtyp2}
E_\gamma L_{E_\gamma}^{\rm IC} 
\propto
\left\{ \begin{array}{ll}
E_\gamma^{(2-q_1)/2}
& \mbox{($E_\gamma\leq E_{\rm IC}^b$)}\\
E_\gamma^{(2-q_2)/2}
& \mbox{($E_{\rm IC}^b<E_\gamma \leq E_{\rm KN,1}$)}\\
E_\gamma^{\beta_{\rm KN}-q_2}
& \mbox{($E_{\rm KN,1}\leq E_\gamma$)}
\end{array} \right.
\end{equation}
where $\beta_{\rm KN}$ reflects the logarithmic energy dependence in the KN cross section.  For example, in the EIC case, one roughly expects $E_\gamma^{\beta_{\rm KN}}\propto \ln[2E_\gamma E_{\rm sn}/(m_e^2c^4)]$. 
For $E_{\rm KN,1}\leq E_{\rm IC}^b$, we have
\begin{equation}\label{eq:EICtyp}
E_\gamma L_{E_\gamma}^{\rm IC} 
\propto
\left\{ \begin{array}{ll} 
E_\gamma^{(2-q_1)/2}
& \mbox{($E_\gamma\leq E_{\rm KN,1}$)}\\
E_\gamma^{\beta_{\rm KN}-q_1}
& \mbox{($E_{\rm KN,1}<E_\gamma\leq E_{\rm KN,2}$)}\\
E_\gamma^{\beta_{\rm KN}-q_2}
& \mbox{($E_{\rm KN,2}\leq E_\gamma$)}
\end{array} \right.
\end{equation}
In the latter case, $E_{\rm KN,1}=E_{\rm KN}^{\rm typ}$ and $E_{\rm KN,2}=E_{\rm KN}^b$.  As seen in Section 3, these spectra are typically anticipated for the generated EIC emission from the early PWN.

While we have provided analytical estimates, as presented in Section~3, we perform numerical calculations to show resulting X-ray and gamma-ray spectra.  This is because not only the KN effect is relevant but also high-energy gamma rays may not escape from the PWN due to the $\gamma\gamma\rightarrow e^+e^-$ process.  
As a result, as shown in Section~3, detailed numerical spectra may deviate from the above analytical estimates even though they come to a reasonable agreement.   
In this work, for the intrinsic emission from the PWN, we solve the following kinetic equations: 
\begin{eqnarray}\label{eq:cascade}
\frac{\pd n_{E_e}^e}{\pd t} &=& \frac{\pd n_{E_e}^{(\gamma \gamma)}}{\pd t} 
- \frac{\pd}{\pd E} [(P_{\rm IC}+P_{\rm syn}+P_{\rm ad}) n_{E_e}^e] 
+ \dot{n}_{E_e}^{\rm inj},\nonumber\\
\frac{\pd n_{E_\gamma}^\gamma}{\pd t} &=& -\frac{n_{E_\gamma}^{\gamma}}{t_{\gamma \gamma}} - \frac{n_{E_\gamma}^{\gamma}}{t_{\rm esc}^{\rm nb}}
+ \frac{\pd n_{E_\gamma}^{(\rm IC)}}{\pd t} 
+ \frac{\pd n_{E_\gamma}^{(\rm syn)}}{\pd t},
\end{eqnarray}
where 
\begin{eqnarray}
t_{\gamma \gamma}^{-1} &=& \int d E_\gamma \,\, n_{E_\gamma}^\gamma  \int \frac{d \cos\theta}{2} \,\, \tilde{c} \sigma_{\gamma \gamma} 
, \nonumber \\
\frac{\pd n_{E_\gamma}^{(\rm IC)}}{\pd t} &=& \int d E_e \,\, n_{E_e}^e \, \int d E_\gamma\,\, n_{E_\gamma}^\gamma \, \int  \frac{d \cos\theta}{2} \,\, \tilde{c}   \frac{d \sigma_{\rm IC}}{d E_\gamma} 
, \nonumber \\
\frac{\pd n_{E_e}^{(\gamma \gamma)}}{\pd t} &=& \frac{1}{2} \int d E_\gamma\,\, n_{E_\gamma}^\gamma \, \int d E'_\gamma\,\, n_{E'_\gamma}^\gamma \, \int \frac{d \cos\theta}{2} \,\, \tilde{c}  \frac{d \sigma_{\gamma \gamma}}{d E_e} 
, \nonumber
\end{eqnarray}
Here $\tilde{c} =(1-\cos\theta)c$ (where $\theta$ is the angle between two particles), $t_{\gamma\gamma}$ is the two-photon annihilation time, $t_{\rm esc}^{\rm nb}=R_w/c$ is the photon escape time for the PWN, $P_{\rm IC}$ is the IC energy-loss rate, $P_{\rm syn}$ is the synchrotron energy-loss rate, and $P_{\rm ad}$ is the adiabatic energy-loss rate~\footnote{A factor of $1/2$ is introduced to avoid double counting.  But, it is unnecessary in the linear-cascade problem, when projectile and target photon spectra are explicitly separated.}.  To save calculation time, we use the continuous energy-loss approximation for the IC process, and assume $E_e=(E_\gamma+E'_\gamma)/2$ for pairs produced by $\gamma\gamma\rightarrow e^+e^-$.  The pair injection rate $\dot{n}_{E_e}^{\rm inj}$ is determined via Equation~(\ref{eq:inj}).  For simplicity, we consider a one-zone model, assuming that only freshly accelerated leptons are relevant.  
We solve the above equations for the constant injection with $\dot{n}_{E_e}^{\rm inj}$.  For the initial conditions, we use $n_{E_e}^e=0$ and $n_{E_\gamma}^\gamma$ is set to a blackbody spectrum with $T_{\rm sn}$.  Resulting high-energy photons are produced by injected non-thermal electrons.  The calculation is performed during the dynamical time $t_{\rm dyn}$, and we essentially obtain steady-state spectra.  Energy boundaries are set to ${10}^{-4}$~eV and ${10}^{16}$~eV with 400 logarithmic energy bins.  
The differential luminosity before attenuation, which is related to ${\mathcal N}_{E_\gamma}$, is calculated by 
\begin{equation}
E_\gamma L_{E_\gamma}= \frac{(E_\gamma^2 n_{E_\gamma}){\mathcal V}_w}{t_{\rm esc}^{\rm nb}},
\end{equation}
which gives observed X-ray and gamma-ray fluxes.

\subsection{Two-photon Annihilation in Embryonic PWNe}
The SN emission and non-thermal synchrotron emission may prevent high-energy gamma rays from leaving the PWN via $\gamma \gamma \rightarrow e^+e^-$.  We also take into account the gamma-ray attenuation (and subsequent regeneration) in the PWN.  The two-photon annihilation cross section is given by
\begin{eqnarray}
\sigma_{\gamma \gamma}&=&\frac{3}{16} \sigma_T (1-\beta_{\rm cm}^2) \left[2 \beta_{\rm cm} (\beta_{\rm cm}^2-2) \right.\nonumber\\
&+& \left.(3-\beta_{\rm cm}^4) \ln[(1+\beta_{\rm cm})/(1-\beta_{\rm cm})] \right], 
\end{eqnarray}
where $\beta_{\rm cm}=\sqrt{(1-4 m_e^2 c^4/S)}$ and $S$ is the Mandelstam variable. 
For a thermal photon spectrum, using the SN photon density $n_\gamma^{\rm sn}=2\zeta(3){(kT_{\rm sn})}^3/(\pi^2\hbar^3c^3)$, the optical depth to pair production is approximated to be
\begin{eqnarray}
\tau_{\gamma \gamma_{\rm sn}}^{\rm ej}&\approx&\frac{3}{16}\sigma_T n_\gamma^{\rm sn} R_{\rm ej}{\mathcal G}\left(x=\frac{m_e^2c^4}{E_\gamma kT_{\rm sn}}\right)\nonumber\\
&\simeq&2.0\times{10}^{4}~{\left(\frac{kT_{\rm sn}}{1~{\rm eV}}\right)}^3{\left(\frac{V_{\rm ej}}{5000~{\rm km}~{\rm s}^{-1}}\right)}t_7,
\end{eqnarray}
where the function ${\mathcal G}(x)\equiv{\mathcal F}(x)/\zeta(3)$ and ${\mathcal F}(x)$ is defined in \cite{der+12}.  In the last expression, we have used ${\mathcal G}(x)\simeq1$ at 
\begin{equation}
E_{\gamma\gamma}^{\rm typ}\approx\frac{m_e^2c^4}{2kT_{\rm sn}}\simeq130~{\rm GeV}~{\left(\frac{kT_{\rm sn}}{1~{\rm eV}}\right)}^{-1}.  
\end{equation}
When non-thermal synchrotron emission provides target photons, for a power-law spectrum with $n_E^{\rm syn}\propto E^{-\beta}$, the optical depth to pair production in the PWN is estimated to be
\begin{eqnarray}
\tau_{\gamma \gamma_{\rm syn}}^{\rm nb}&\approx&0.2\sigma_T (E n_E^{\rm syn}) R_{w}\nonumber\\
&\simeq&3.1\times{10}^{-3}~\gamma_{b,5}^{-2}P_{i,-2.5}B_{\rm dip,14}^{-2}\epsilon_{B,-2}^{-1/2}\nonumber \\
&\times&{\left(\frac{V_{\rm ej}}{5000~{\rm km}~{\rm s}^{-1}}\right)}^{1/2}\frac{t_7^{-3/2}}{(1+Y)}{\left(\frac{E_\gamma}{E_{\gamma\gamma}^{\rm typ}}\right)}^{\beta-1},
\end{eqnarray}
where $R_w\approx R_{\rm ej}=V_{\rm ej}t$ is used for analytical estimates here.  The typical energy $E_{\gamma\gamma}^{\rm typ}$ is given by
\begin{eqnarray}
E_{\gamma\gamma}^{\rm typ}\approx\frac{m_e^2c^4}{E_{\rm syn}^b}&\simeq&4.1\times{10}^{-2}~{\rm GeV}~\gamma_{b,5}^{-2}P_{i,-2.5}\epsilon_{B,-2}^{-1/2}\nonumber\\
&\times&{\left(\frac{V_{\rm ej}}{5000~{\rm km}~{\rm s}^{-1}}\right)}^{3/2}t_7^{3/2}. 
\end{eqnarray}
In this work, electromagnetic cascades are calculated for emission generated in the PWN.  If $R_w<R_{\rm ej}$, we take into account the further attenuation by SN photon fields, by multiplying ${\rm e}^{-(\tau_{\gamma\gamma}^{\rm ej}-\tau_{\gamma\gamma}^{\rm nb})}$.    

In Figure~3, we show the optical depth to the two-photon annihilation at different times.  One sees that SN photons at optical or infrared bands prevent TeV gamma rays from leaving the PWN.  TeV gamma rays are expected to escape from the source in a few years.  On the other hand, escape of GeV gamma rays is much more promising but even GeV gamma rays can be strongly attenuated by synchrotron photons for $t\lesssim30$~d.  

\begin{figure}
\includegraphics[width=\linewidth]{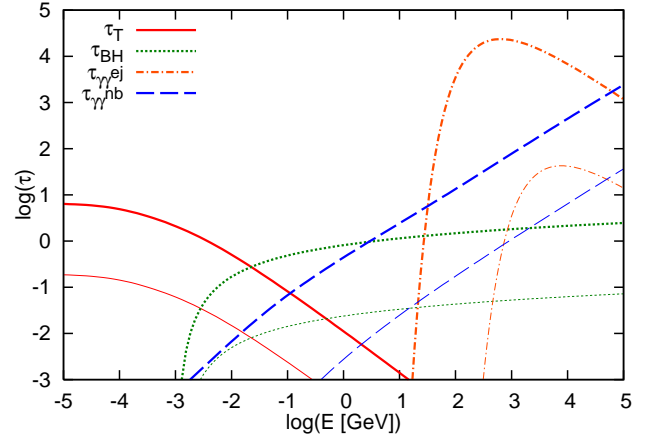}
\caption{Optical depths to Compton (solid curves), Bethe-Heitler (dotted curves), and two-photon annihilation (dot-dashed curves for thermal targets; dashed curves for non-thermal targets) processes, for ($P_{i}$, $B_{\rm dip}$, $M_{\rm ej}$)=(2~ms, ${10}^{14}$~G, 5~$M_{\odot}$).  Thick and thin curves are for $t={10}^{6.75}$~s and $t={10}^{7.5}$~s, respectively.  One sees that not only SN photons but also synchrotron photons are relevant for two-photon annihilation.  Synchrotron photons can prevent multi-GeV gamma rays from leaving the emission.  
}
\end{figure}

\subsection{Matter Attenuation in the Stellar Material}
Photons escaping from the PWN can be significantly attenuated in the SN ejecta.  Although we avoid detailed radiative-transfer calculations, we approximately account for it as a post-process.  For photons with energies below $\sim10-30$~keV, the most important process is photoelectric absorption.  In the soft X-ray band, ionization breakout emission provides an interesting signal from millisecond pulsars embedded in the SN ejecta~\citep{met+14}.  In this work, we are interested in hard X rays and gamma rays, which are produced via non-thermal processes, so we mainly focus on Compton scattering and Bethe-Heitler (BH) pair production, which are dominant at high energies.  The optical depth is given by $\tau=\tau_{\rm pe}+\tau_{\rm comp}+\tau_{\rm BH}$, where $\tau_{\rm pe}$ is the photoelectric absorption optical depth.  Using the mass attenuation coefficient $K$, it is generally expressed to be $\tau\approx K\rho R$, where $\rho$ is the density and $R$ is the size.  
The photoelectric absorption at high energies is taken into account, using the bound-free opacity $K_{\rm bf}\simeq2.37~{\rm cm}^2~{\rm g}^{-1}~{(Z/6)}^3{(E_{\gamma}/10~{\rm keV})}^{-3}$ for conservative estimates of X-ray emission.  

\begin{figure}
\includegraphics[width=\linewidth]{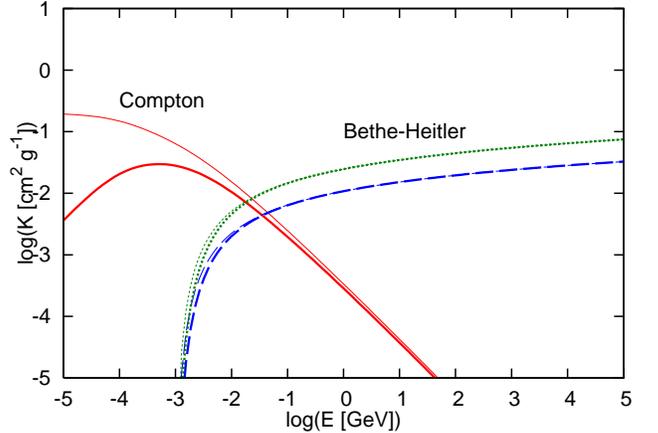}
\caption{Mass attenuation (thin curves) and mass energy-transfer (thick curves) coefficients as a function of photon energy $E$.  Dashed and dotted curves are for $Z_{\rm eff}=2.5$ and $Z_{\rm eff}=7$, respectively.   
}
\end{figure}

The Compton optical depth in the ejecta is 
\begin{equation}
\tau_{\rm comp}^{\rm ej}\approx K_{\rm comp}\rho_{\rm ej}R_{\rm ej}=\frac{(3-\delta)M_{\rm ej}\sigma_{\rm comp}}{4\pi\mu_em_uR_{\rm ej}^2},
\end{equation}
which is reduced to Equation (\ref{eq:Thomson}) at low energies of $E_\gamma\lesssim10$~keV, and $K_{\rm comp}=\sigma_{\rm comp}/(\mu_e m_u)$.  The mass energy-transfer coefficient is obtained using
\begin{eqnarray}
\kappa_{\rm comp} \sigma_{\rm comp}&=&\frac{3}{4}\sigma_T\left[\frac{2{(1+x)}^2}{x^2(1+2x)}-\frac{1+3x}{{(1+2x)}^2}\right.\nonumber\\
&-&\frac{(1+x)(2x^2-2x-1)}{x^2{(1+2x)}^2}-\frac{4x^2}{3{(1+2x)}^3}\nonumber\\
&-&\left.\left(\frac{1+x}{x^3}-\frac{1}{2x}+\frac{1}{2x^3}\right)\ln(1+2x) \right],
\end{eqnarray}
which is obtained from the known KN cross section and kinematics.  Here $x\equiv E_\gamma/(m_ec^2)$ and $\kappa_{\rm comp}$ is the gamma-ray inelasticity. 

At high energies, the BH pair production process is dominant.  For a nucleus with mass number $A$ and atomic number $Z$, the BH process on a nucleus scales as $\sigma_{\rm BH}=Z^2\sigma_{\rm BH}^{(p)}$.  
Taking into account contributions from both nuclei and electrons, for $\mu_e\approx2$, we have
\begin{equation}
\tau_{\rm BH}^{\rm ej}\approx\frac{(3-\delta)M_{\rm ej}(Z_{\rm eff}+1)\sigma_{\rm BH}^{(p)}}{8\pi m_uR_{\rm ej}^2},  
\end{equation}
where $Z_{\rm eff}$ is the effective atomic number, which depends on chemical composition of the ejecta.  For $X_{\rm H}=0.6$ and $X_{\rm He}=0.3$ and $X_{\rm C}=0.1$, we obtain $Z_{\rm eff}\approx2.5$, while we may have $Z_{\rm eff}\approx7$ for $X_{\rm CO}=1$.  The mass energy-transfer coefficient at high energies is approximately obtained from
\begin{equation}
\kappa_{\rm BH} \sigma_{\rm BH}=\frac{x-2}{x}\sigma_{\rm BH},
\end{equation}
neglecting contributions from electron-positron annihilation.  In this work, we use the cross section derived from the Born approximation~\citep{cho+92}.  For analytical estimates, one may use a simpler formula
\begin{equation}
\sigma_{\rm BH}^{(p)}\approx \frac{3\alpha}{8\pi}\sigma_T\left[\frac{28}{9}\ln(2x)-\frac{218}{27}\right],
\end{equation}
which gives $\sigma_{\rm BH}\sim Z^2{10}^{-26}~{\rm cm}^2$ at GeV energies.  Note that the BH cross section is the order of $\sigma_{\rm BH}^{(p)}\sim \alpha_{\rm em}\sigma_T$.  At GeV energies, the BH optical depth is estimated by
\begin{eqnarray}\label{eq:BH}
\tau_{\rm BH}^{\rm ej}&\simeq&0.57~([Z_{\rm eff}+1]/3)(2/\mu_e)(M_{\rm ej}/5~M_\odot)\nonumber\\
&\times&{(V_{\rm ej}/5000~{\rm km}~{\rm s}^{-1})}^{-2}t_7^{-2},
\end{eqnarray}
implying that the BH attenuation is significant at very early times.  But, since the BH cross section is $\alpha_{\rm em}\simeq1/137$ times lower than the Thomson cross section, GeV gamma rays allow us to probe subphotospheric regions, i.e., optically-thick phases such that $\tau_T^{\rm ej}\gtrsim1$.    
 
In the small inelasticity limit, a particle loses $\kappa_\gamma$ per interaction, so the survival fraction is ${(1-\kappa_\gamma)}^{\rm max[\tau,\tau^2]}$, where $\rm max[\tau,\tau^2]$ represents the number of scattering.  In the large inelasticity limit, as in the attenuation case, the survival fraction is given by ${\rm e}^{-\tau}$.    
Combing the two limits, we approximate the escape fraction of hard X rays and gamma rays by 
\begin{equation}\label{eq:scat}
f_{\rm esc} = {\rm e}^{-\tau}+(1-{\rm e}^{-\tau})(1-\kappa)^{\rm max[\tau,\tau^2]}. 
\end{equation}

In Figure~4, we show mass attenuation and mass energy-transfer coefficients.  For $M_{\rm ej}=10~M_\odot$, we use $Z_{\rm eff}=2.5$ assuming a typical composition for ejecta of Type II SNe, whereas we use $Z_{\rm eff}=7$ for $M_{\rm ej}=5~M_\odot$ assuming ejecta are dominated by carbon and oxygen.  Different chemical compositions lead to modest influences on attenuated gamma-ray spectra, but our conclusions are not qualitatively altered.  In Figure~3, optical depths to Compton and BH pair production processes are shown, respectively.  Obviously, GeV-TeV gamma rays cannot leave the ejecta until a few months after the explosion.  The GeV gamma-ray escape is allowed at
\begin{eqnarray}
t_{\gamma-\rm bo}&\simeq&88~{\rm d}~{([Z_{\rm eff}+1]/3)}^{1/2}{(M_{\rm ej}/5~M_\odot)}^{1/2}\nonumber\\
&\times&{(V_{\rm ej}/5000~{\rm km}~{\rm s}^{-1})}^{-1}.
\end{eqnarray}
In the Thomson limit, the gamma-ray flux at $E_{\rm IC}^b$ for $t\gtrsim t_{\rm em}$ is roughly given by
\begin{eqnarray}
F_{\rm IC}^b&\sim&3.7\times{10}^{-8}~{\rm GeV}~{\rm cm}^{-2}~{\rm s}^{-1}~Y{(1+Y)}^{-1}B_{\rm dip,14}^{-2}\nonumber\\
&\times&{([Z_{\rm eff}+1]/3)}^{-1}{(M_{\rm ej}/5~M_\odot)}^{-1}{(V_{\rm ej}/5000~{\rm km}~{\rm s}^{-1})}^{2}\nonumber\\
&\times&{(d/16.5~{\rm Mpc})}^{-2}{(t/t_{\gamma-\rm bo})}^{-2}. 
\end{eqnarray}
Note that the ejecta becomes optically thin to Thomson scattering at
\begin{eqnarray}
t_{\rm HX-bo}&\simeq&420~{\rm d}~{(2/\mu_e)}^{1/2}{(M_{\rm ej}/5~M_\odot)}^{1/2}\nonumber\\
&\times&{(V_{\rm ej}/5000~{\rm km}~{\rm s}^{-1})}^{-1}.
\end{eqnarray}  
The synchrotron flux at late times is estimated to be
\begin{eqnarray}
F_{\rm syn}^b&\sim&2.6\times{10}^{-12}~{\rm erg}~{\rm cm}^{-2}~{\rm s}^{-1}~{(1+Y)}^{-1}B_{\rm dip,14}^{-2}\nonumber\\
&\times&{(2/\mu_e)}^{-1}{(M_{\rm ej}/5~M_\odot)}^{-1}{(V_{\rm ej}/5000~{\rm km}~{\rm s}^{-1})}^{2}\nonumber\\
&\times&{(d/16.5~{\rm Mpc})}^{-2}{(t/t_{\rm HX-bo})}^{-2}. 
\end{eqnarray}
Note that low-energy photons with low $\kappa_\gamma$ can escape earlier after they experience multiple scatterings.

\section{Numerical Results}
We solve Equations~(\ref{eq:L}), (\ref{eq:nb}), (\ref{eq:ej}), (\ref{eq:int}), (\ref{eq:th}) and (\ref{eq:cascade}) numerically.  Then, we approximately take into account matter attenuation in the SN ejecta via Equation~(\ref{eq:scat}).  
Key parameters for dynamics are $P_i$, $B_{\rm dip}$, $M_{\rm ej}$ and ${\mathcal E}_{\rm sn}$.  Throughout this work, ${\mathcal E}_{\rm sn}=2\times{10}^{51}$~erg is used.  We also consider $M_{\rm ej}=5~M_{\odot}$ and $M_{\rm ej}=10~M_{\odot}$, which are often suggested from modeling of observed Type Ibc and II SNe, respectively.  To detect X-ray and gamma-ray emission, sufficiently fast-rotating and magnetized NSs are required, so we consider NSs with $P_i\leq10$~ms and $B_{\rm dip}\geq{10}^{13}$~G.  Other microphysical parameters are treated as sub-parameters, assuming that they are similar to ones suggested in the literature of Galactic PWNe.  Motivated by results on the Crab nebula~\citep{tt10}, we assume $\epsilon_B=0.003$ and $\epsilon_e=1-\epsilon_B$, fixing $q_1=1.5$, $q_2=2.5$ and $\gamma_b={10}^{5.5}$.  

\begin{figure}
\includegraphics[width=\linewidth]{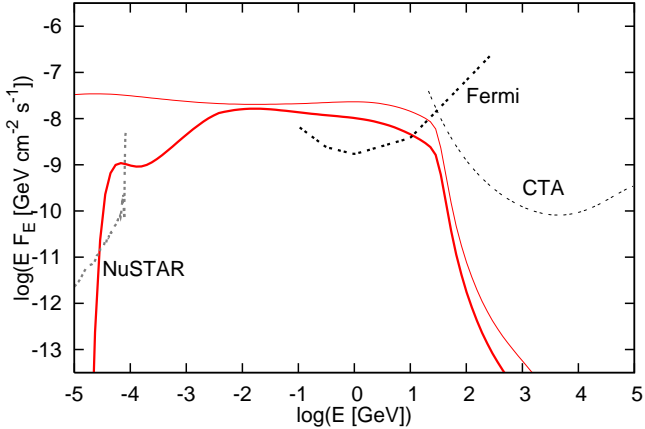}
\caption{High-energy photon spectra of the early PWN embedded in the SN ejecta.  The observation time is set to $t={10}^{6.75}~{\rm s}\simeq65$~d, and the source distance is taken as $d=16.5$~Mpc.  Relevant parameters for dynamics are ($P_{i}$, $B_{\rm dip}$, $M_{\rm ej}$)=(2~ms, ${10}^{14}$~G, 5~$M_{\odot}$). 
We show cases with (thick curve) and without (thin curve) matter attenuation. Note that cascades via $\gamma\gamma\rightarrow e^+e^-$ in the emission region is considered. The {\it Fermi}/LAT sensitivity at the corresponding observation time and {\it NuSTAR} (${10}^6$~s) and CTA (50~hr) sensitivities~\citep{cta+11} are also overlaid. 
}
\end{figure}
\begin{figure}
\includegraphics[width=\linewidth]{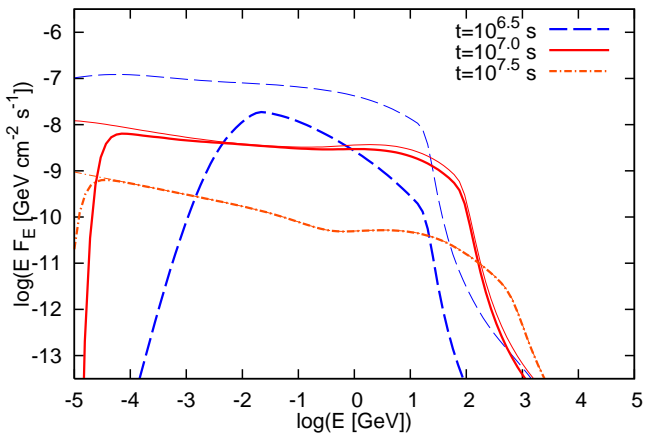}
\caption{The same as Figure~5, but at different observation times.
}
\end{figure}
\begin{figure}
\includegraphics[width=\linewidth]{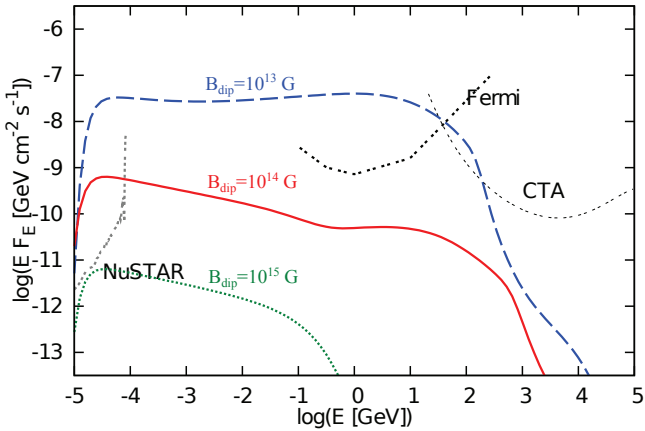}
\caption{High-energy photon spectra of the early PWN embedded in the SN ejecta for $P_i=2$~ms at $t={10}^{7.5}~{\rm s}\simeq316$~d.  Different magnetic field strengths are considered.  Detections with CTA are possible for $B_{\rm dip}={10}^{13}$~G.  
}
\end{figure}
\begin{figure}
\includegraphics[width=\linewidth]{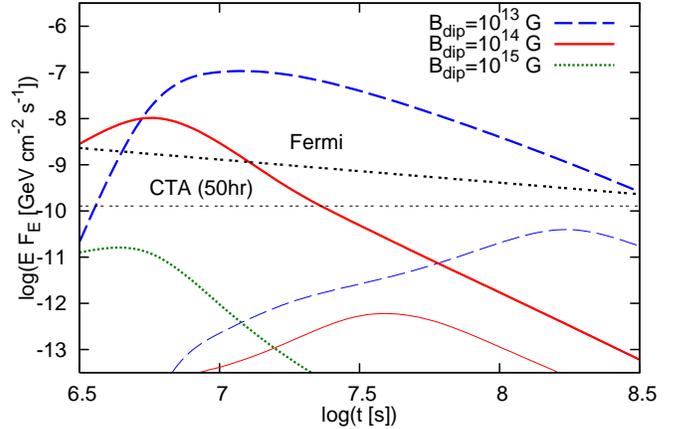}
\caption{High-energy gamma-ray light curves of the early PWN emission at 1~GeV (thick curves) and 1~TeV (thin curves), for different magnetic field strengths.  The {\it Fermi}/LAT and CTA sensitivities are overlaid.  Note that the observation time 50~hr is shorter than $t$ for CTA, while we consider continuous observation for {\it Fermi}/LAT so the sensitivity changes as $t^{-1/2}$.    
}
\end{figure}

In Figure~5, we show hard X-ray and gamma-ray spectra for the millisecond pulsar case with ($P_{i}$, $B_{\rm dip}$, $M_{\rm ej}$)=(2~ms, ${10}^{14}$~G, 5~$M_{\odot}$).  The observation time is set to $t={10}^{6.75}$~s after the explosion.  As expected in Equation~(\ref{eq:EICtyp}), generated EIC emission has a peak around $\sim10-100$~GeV.  Although our numerical results and analytical estimates (presented in Section~2.2) come to a reasonable agreement, detailed effects due to energy-dependent cross sections and electromagnetic cascades play roles in making difference.  Two-photon attenuation is still important at early times.  One sees that the spectrum below $\sim3$~GeV is also softened due to interactions with synchrotron photons (see~Figure~3), and that there is a prominent cutoff at $\sim30$~GeV, due to SN photons.  Furthermore, matter attenuation makes the gamma-ray spectrum even softer.  Time evolution of spectra is also shown in Figure~6.  Not only various attenuation processes but also the KN effect becomes less important as time, and the generated EIC spectrum at $\sim1$~yr is described rather by Equation~(\ref{eq:EICtyp2}).  At late times, SN emission becomes so weak that the EIC emission is less important.  As a result, the synchrotron component is more prominent, and one sees the synchrotron cutoff expected by Equation~(\ref{eq:syncut}).   

We find that $\sim1-10$~GeV gamma rays can be detected by {\it Fermi} several months after the SN explosion.  Here we consider nearby SNe at $d=16.5$~Mpc, motivated by the possibility that GWs from newborn fast-rotating NSs in the Virgo cluster can be detected by second-generation ground-based GW interferometers~\citep[][]{ste+05,dal+09}.  Hard X-ray observations by high-sensitivity satellites such as {\it NuSTAR} look more promising although followup observations are required.  As suggested in Figure~7, such pulsar-powered SNe may be detected up to $d\sim0.1-1$~Gpc, depending on values of $B_{\rm dip}$.  

Note that transients like GRBs and SNe have been potentially interesting targets for imaging atmospheric Cherenkov telescopes and CTA can be especially powerful for that purpose~\citep{kak+12,ino+13,bar+14}.  However, for embryonic PWNe, TeV gamma-ray detections may be challenging due to limitation of the maximum energy (see Equation~\ref{eq:ICmax}), the KN effect, and two-photon attenuation, although CTA might be able to detect the signal at late times for appropriate values of $B_{\rm dip}$ (see Figure~7).  
In Figure~8, we show gamma-ray light curves for different magnetic field strengths.  For $P_i\sim1-3$~ms, it is difficult for {\it Fermi} to detect GeV gamma rays for magnetic fields with $B_{\rm dip}\gtrsim{10}^{14.5}$~G, because the spin-down power rapidly declines at $t\gg t_{\rm em}$ and the target SN photon density also decreases as time.  However, hard X rays are still detectable even for such newborn magnetars, because the synchrotron component decays as $t^{-2}$ and sensitivities of followup X-ray observations are better than that of {\it Fermi}.  For a nearby SN at the Virgo cluster, GeV gamma-ray detections are feasible for magnetic fields down to $B_{\rm dip}\lesssim{10}^{12}$~G.     

\begin{figure}
\includegraphics[width=\linewidth]{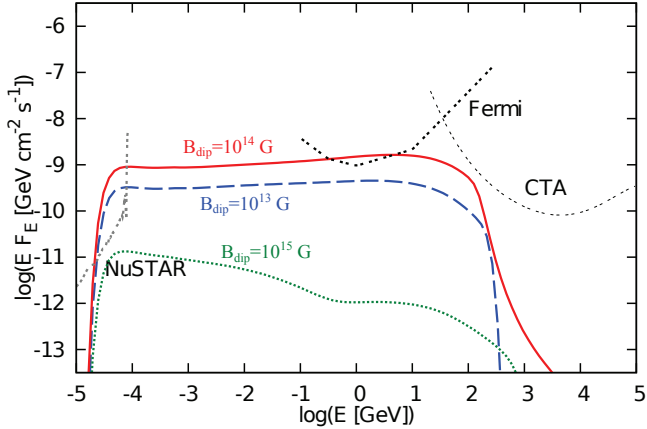}
\caption{High-energy photon spectra of the early PWN embedded in the SN ejecta for $P_i=10$~ms at $t={10}^{7.25}~{\rm s}\simeq206$~d.  Different magnetic field strengths are considered.
}
\end{figure}
\begin{figure}
\includegraphics[width=\linewidth]{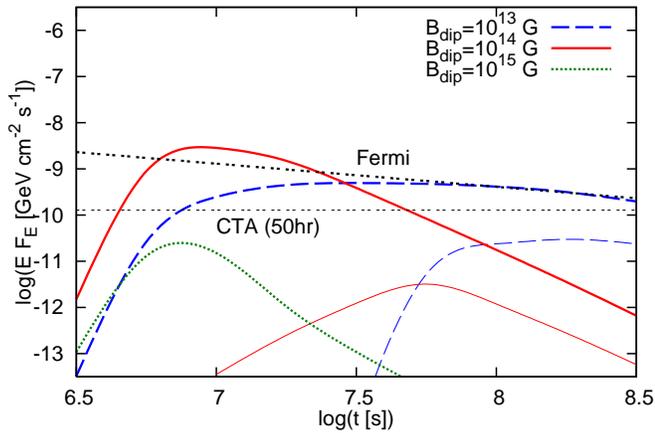}
\caption{The same as Figure~8, but for $P_i=10$~ms. 
}
\end{figure}

Newborn millisecond pulsars have been postulated to explain energetic transients such as super-luminous SNe, hypernovae and GRBs, but it is not clear how newborn NSs acquire such fast rotation.  Hence, we consider a more conservative case of $P_i=10$~ms, which is not far from values inferred for the Crab pulsar and PSR J0537-6910~\citep[][]{fk06}.  Here, the rotational energy is smaller than the SN explosion energy, so SN dynamics is essentially unaffected by the spin-down power.  In Figures~9 and 10, we show spectra and light curves, respectively.  Interestingly, we may still expect that GeV gamma rays can be detected for a SN at the Virgo cluster for ${10}^{13}~{\rm G}\lesssim B_{\rm dip}\lesssim{10}^{14}$~G.  Hard X rays are more promising since the synchrotron signal can be seen up to $d\sim50-100$~Mpc for $B_{\rm dip}\sim{10}^{13}-{10}^{14}$~G, although detections become challenging for sufficiently strong magnetic fields with $B_{\rm dip}\gtrsim{10}^{14.5}$~G.  

The core-collapse SN rates within 20~Mpc and within 50~Mpc are estimated to be $\sim3~{\rm yr}^{-1}$ and $\sim50~{\rm yr}^{-1}$, respectively.  However, we expect that only a fraction of SNe can leave NSs with fast rotation.  The most optimistic case is motivated by the dynamo hypothesis for magnetars, which requires fast rotation.  Since the magnetar fraction is believed to be $\sim10$\% of all NSs, the probability to expect high-energy counterparts may not be so low.  Hypernovae or trans-relativistic SNe associated with low-luminosity GRBs, which are often thought to be engine-driven SNe, could come from fast-rotating pulsars, and their rates are typically a few percent of the core-collapse SN rate.  If we assume that 2\% of SNe lead to such SNe, their rate within 50~Mpc is estimated to be $\sim1~{\rm yr}^{-1}$, which is encouraging.  For GeV gamma rays, we suggest that individual searches and stacking analyses for super-luminous SNe and hypernovae may be more promising.  In particular, in view of modeling of light curves and energetics, hydrogen-poor super-luminous SNe are interesting targets~\citep{qui+11,vre+14}, and finding evidence for gamma rays from these SNe can support the hypothesis that they are driven by newborn pulsars.  

In Figures~11 and 12, we show the dependence on $M_{\rm ej}$.  Obviously, for the same values of $B_{\rm dip}$ and $P_i$, escape of photons is more difficult for larger $M_{\rm ej}$.  However, after hard X rays or gamma rays break out (i.e., $\tau_T^{\rm ej}\lesssim1$ or $\tau_{\rm BH}^{\rm ej}\lesssim1$), one sees that the dependence on $M_{\rm ej}$ is quite modest.  This also implies that the most important parameters in our model are $B_{\rm dip}$ and $P_i$.  
\begin{figure}
\includegraphics[width=\linewidth]{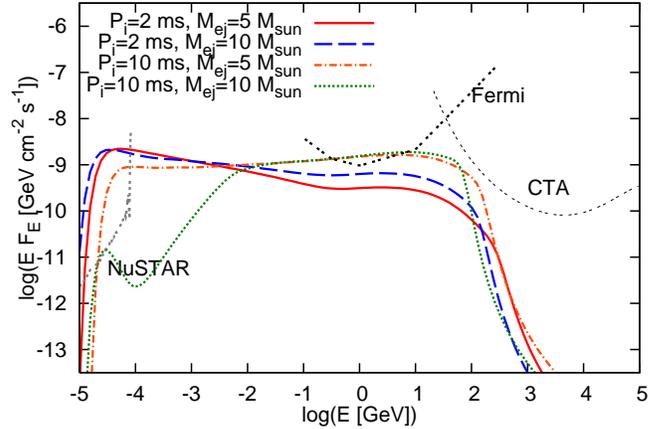}
\caption{High-energy photon spectra of the early PWN embedded in the SN ejecta for different values of $P_i$ and $M_{\rm ej}$.  The observation time is set to $t={10}^{7.25}~{\rm s}\simeq206$~d. 
}
\end{figure}
\begin{figure}
\includegraphics[width=\linewidth]{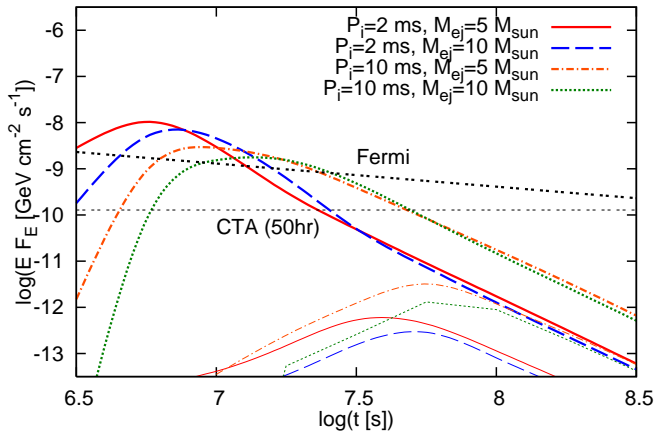}
\caption{The same as Figure~8, but for different values of $P_i$ and $M_{\rm ej}$. 
}
\end{figure}

\section{Discussion and Summary}
We showed that high-energy emission from embryonic PWNe provides a useful probe of particle acceleration at very early stages of SNe leaving a fast-rotating NS remnant.  We described analytical spectra of generated IC emission, taking into account the KN effect that is important above $\sim10-100$~GeV.  Initially, two-photon attenuation prevents gamma rays from leaving the PWN and further attenuation in the SN ejecta is unavoidable.  Although analytical estimates and numerical calculations reach a reasonable agreement, details of microphysical processes, including electromagnetic cascades, affect the resulting gamma-ray spectra.   After several months, the PWN itself becomes transparent to gamma rays, and GeV gamma rays break out before optical photons and X rays escape.  

In this work, we focused on presenting detailed spectra and light curves.  We will carry out a parameter survey on the detectability in the plane of $B_{\rm dip}$, $P_i$ and $d$ in future.  Nevertheless, here we briefly discuss detectability in a typical scenario to provide a picture on the prospects of current and future observations.  If such a SN in a galaxy in the Virgo cluster leaves a NS with $P_i=2$~ms, gamma rays can be detected by {\it Fermi} for ${10}^{12}~{\rm G}\lesssim B_{\rm dip}\lesssim{10}^{14.5}$~G.  For $P_i=10$~ms, detections are possible in narrower ranges ${10}^{13}~{\rm G}\lesssim B_{\rm dip}\lesssim{10}^{14}$~G, and become more difficult for larger values of $P_i$.  Even though detections of IC emission from a single source with GeV gamma rays are limited to nearby sources and the PWN emission can be seen by {\it Fermi} up to $d\lesssim40$~Mpc for $B_{\rm dip}\sim{10}^{14}$~G, we found that gamma rays can still provide us with useful counterparts of GW emission from newborn NSs.  In addition, to test the possibility that super-luminous SNe are driven by fast-rotating NSs~\citep[cf.][]{mur+11isn}, individual and stacking analyses on transients with months-to-years time scales can also be useful.  
We also showed that observations of non-thermal synchrotron emission at hard X rays can be more powerful to identify pulsar-aided SNe, although dedicated followup observations (e.g., by {\it NuSTAR}) are required.  A newborn NS with $P_i=2$~ms at $d=16.5$~Mpc can be detected for reasonable magnetic fields ${10}^{11.5}~{\rm G}\lesssim B_{\rm dip}\lesssim{10}^{15}$~G with observation time $\sim{10}^5-{10}^6$~s.  For $P_i=10$~ms, ${10}^{12}~{\rm G}\lesssim B_{\rm dip}\lesssim{10}^{14.5}$~G is needed, and {\it NuSTAR}-like detectors can detect a source up to $d\sim80$~Mpc for $B_{\rm dip}\sim{10}^{14}$~G.  It is likely that only a fraction of core-collapse SNe can leave NSs with $P_i\lesssim10$~ms.  Assuming that a fraction $f_{r}$ of NSs can be rapidly rotating and we can detect all SNe in the nearby universe with surveys such as ASAS-SN~\footnote{http://www.astronomy.ohio-state.edu/~assassin/index.shtml}, the rate to have such events within $d$ may be $\sim0.5~{\rm yr}^{-1}f_{r,-2}{(d/50~{\rm Mpc})}^{3}$, which is in the interesting range.  

One may expect not only high-energy gamma rays but also high-energy neutrinos as GW counterparts.  Neutrino detections from newborn NSs are also promising for nearby SNe.  Until the proto-PNS becomes transparent to neutrinos, baryons including neutrons may be loaded into the proto-NS wind via MeV neutrinos.  Neutrons that are initially coupled to ions should be magnetically accelerated together.  Then, dissipation of relativistic neutron flows inevitably lead to GeV-TeV neutrino production.  Since the baryon loading is not small at early stages, ion acceleration before and/or at the termination shock could also be efficient, where TeV or higher-energy neutrino production is possible~\citep[][]{mur+14}.  At later times after the proto-PNS becomes transparent to neutrinos (i.e., $t\gg10-100$~s), it would be more natural that the wind is largely dominated by electrons and positrons.  But, it has also been speculated that dissipation in the current sheet may lead to ion acceleration and observed cosmic rays can be explained~\citep[][]{og69,aro03,fan+13}.  If this is the case, EeV neutrinos provide a powerful test of this hypothesis~\citep{mur+09,fan+14}. 

Our assumption is that phenomenology of Galactic PWNe can be extrapolated to early PWNe.  The detectability is sensitive to $B_{\rm dip}$ and $P_i$ and it is theoretically unclear how such fast-spinning NSs are born.  Indeed, although hard X-ray observations at early times are more relevant for sufficiently large $B_{\rm dip}$, non-observations in the soft X-ray range suggest that only fraction of NSs can be born with fast rotation~\citep[][]{per+08}.  There is no evidence for such a pulsar in SN 1987A.  On the other hand, high-energy emission is detectable even for non-extreme values of $P_i$ and $B_{\rm dip}$ as in the Crab pulsar, and possible candidates have been reported for historical SNe in X rays~\citep[][]{sp08}.  We encourage individual and stacking searches using {\it Fermi} data.  Although the high fraction of such events has apparently been constrained by non-observations, some could be unidentified transients.  Importantly, successful detections should allow us to study the beginning of particle acceleration in PWNe.  Magnetic dissipation and subsequent particle acceleration mechanisms have been long-standing problems~\citep[see, e.g.,][and references therein]{hos+92,kir10,ss11,aro12}.  High-energy signals would infer that the Poynting energy is efficiently converted into the particle energy, and one can put constraints on $\epsilon_B$ by observing the IC component.  Also, if possible, measurements of the spectral shape would also be useful for constraining $\eta$, $\gamma_b$ (or $\mu_\pm$) and investigating physical connection to Galactic PWNe.  

If strong non-thermal signatures of embryonic PWNe are detected for super-luminous SNe or hypernovae, it can support that fast-rotating pulsars play a role in SN emission or even its dynamics.  Note that there are several competing scenarios for super-luminous SNe, including the interaction-powered SN scenario and pair-instability SN scenario.  Interaction-powered SNe have been suggested to be sources of hadronic GeV-TeV gamma rays and TeV-PeV neutrinos~\citep[][]{mur+11isn}.  Properties of non-thermal spectra should be different, so discrimination between the two scenarios is possible.  In the interaction-powered SN scenario, non-thermal gamma-ray and neutrino emission becomes prominent around shock breakout from optically-thick circumstellar material (although the reverse-shock neutrino emission could be expected before photons escape), and thermal X-ray and narrow width hydrogen line emission is accompanied~\citep[e.g.,][]{ofe+13,mar+14,ofe+14}.  Although non-thermal hard X rays are expected as well~\citep[][]{mur+11isn}, after the shock breakout and GeV-TeV gamma rays escape, their flux is lower than the gamma-ray flux.  On the other hand, in the pulsar-aided SN scenario, synchrotron X rays are stronger than IC gamma rays.  We suggest that, if pulsars are embedded in the ejecta, high-energy emission is promising especially for hydrogen-poor SNe (although it may also be caused by collisions with circumstellar material)~\citep[][]{qui+11,ben+14,nic+14,che+14,vre+14}.  Also, for normal luminosity SNe, the ratio of TeV emission to GeV emission is lower than that in the interaction-powered SN scenario where gamma rays are hadronically produced.  Although GRBs are not considered in this work, magnetic dissipation and particle acceleration might occur even when the proto-NS wind forms Poynting-dominated jet-like outflows.  We note that shallow-decay afterglow emission is often explained by energy injection via the magnetar spin down~\citep[e.g.,][]{met+11}, and that the observed X-ray emission, which could be associated with the GW emission~\citep[][]{cm09}, can be attributed to synchrotron emission from leptons accelerated via internal magnetic dissipation~\citep[][]{ghi+07,mur+11grb,met+11}.  We also suggest that magnetic dissipation in embryonic PWNe may be relevant for the long-lasting X-ray emission~\citep[][]{mur+11grb}.    

Our calculations can also be applied to high-energy counterparts of short GRBs and double NS mergers.  A pulsar-aided mechanism could also be relevant in double NS mergers given that the equation of state is quite stiff~\citep[][]{kis+14}, where GeV-TeV gamma rays and X rays from embryonic PWNe can be useful as a signature of this scenario. 

\begin{acknowledgements}
K. M. thanks Omer Bromberg, Kunihito Ioka, Boaz Katz, and Eran Ofek for useful discussions.
This work is supported by NASA through Hubble Fellowship Grant No. 51310.01, awarded by the STScI, which is operated by the Association of Universities for Research in Astronomy, Inc., for NASA, under Contract No. NAS 5-26555 (K. M.), Einstein Postdoctoral Fellowship Grant No. PF4-150123
awarded by the Chandra X-ray Center, which is operated by the Smithsonian Astrophysical Observatory, for NASA, under contract NAS 8-03060 (K. K.), and Grant-in-Aid for Scientific Research (25103510, 25105508, 24244028, 24740163) and by HPCI Strategic Program of Japanese MEXT (hp130025, 140211) (K. K.).
\end{acknowledgements}


\bibliography{ref}

\end{document}